\documentclass[preprint,aps,showpacs,preprintnumbers,amsmath,amssymb,footinbib]{revtex4-2}
\usepackage{mathrsfs}
\usepackage{amssymb}
\usepackage{graphicx}
\usepackage{hyperref}
\usepackage{ulem}
\usepackage{tabularx}
\usepackage{array}
\usepackage{color}

\begin{document}

\title{Lorentz-Covariant Spin Operator for spin $1/2$ Massive Fields As a Physical Observable}

\author{Taeseung Choi}
 \email{tschoi@swu.ac.kr}
\affiliation{Division of General Education, Seoul Women's University, Seoul 139-774, Korea}
\affiliation{School of Computational Sciences, Korea Institute for Advanced Study, Seoul 130-012, Korea}
 \author{Yeong Deok Han}
 \email{ydhan2@hanmail.net}
\affiliation{Department of Computer Science and Engineering, Woosuk University, Jincheon, Chungbuk, 27841, Korea}

\begin{abstract}

We derive a relativistic-covariant spin operator for massive case directly from space-time
symmetry in Minkowski space-time and investigate the physical properties of a derived spin operator. In the derivation we require only two conditions: First, a spin operator should be the generator of the $SU(2)$ little group of the Poincar\'e group. Second, a spin operator should covariantly transform under the Lorentz transformation. A space inversion transformation is shown to play a role to derive a unique relativistic-covariant spin operator, we call the field spin operator, whose eigenvalue labels the spin of a massive (classical) field that provides the irreducible representation space of the Poincar\'e group. 
The field spin becomes the covariant spin in the covariant Dirac representation, which is shown to be the only spin that describes the Wigner rotation properly in the covariant Dirac representation. Surprisingly, the field spin also gives the non-covariant spin, which is the FW spin for the positive energy state. We also show that the field spin operator is the unique spin operator that generate the (internal) $SU(2)$ little group transformation of the Poincar\'e group properly.

\end{abstract}


\maketitle

 \section{ Introduction}
The role of the spin of a massive relativistic particle is ever-growing in a wide range of recently developing fields such as quantum information and vortex physics \cite{TernoRMP,Lee22,OurVortex,SilenkoPRA20}. According to the development of spin-based technologies, fundamental understanding of a spin in a relativistic regime becomes essential. Originally, finding spin operators of massive relativistic particles has been a long-standing problem from the
beginning of relativistic quantum mechanics \cite{Dirac,Pryce48,NW,FW,Fradkin,Chakrabarti,Hilgevoord,Fleming64,Berg,Gursey,Bogolubov,RyderGen,Choi13,Brukner19}. However, there is a lack of consensus on the definition of the spin of a massive relativistic particle in relativistic quantum mechanics \cite{Terno2016,Taillebois}. 
 
Bauke {\textit et al.} pointed out that at least seven spin operators for massive relativistic particles have been suggested \cite{Bauke}. They argued that only two candidates, the Foldy-Wouthuysen (FW) and the Pryce spin operator, qualify as proper relativistic spin operators by studying the two conditions: (i) Spin operator should commute with the free particle Dirac Hamiltonian; (ii) Spin operator should satisfy the $su(2)$ algebra and have two eigenvalues $\pm 1/2$. The condition (ii) derives the spatial vector condition of the spin operator through the total angular momentum that is the sum of the orbital and the spin angular momentum. One can easily check that the Pryce spin operator in Bauke {\textit et al.}'s \cite{Bauke} is not reduced to the non-relativistic spin for negative energy states when the momentum approaches to the rest momentum, but shows the dependence on the direction of the momentum. This fact indicates that there is a preferred direction for the Pryce spin operator at {the rest frame of the momentum}, which is not physically sound. Hence the FW spin operator seems to be the proper relativistic spin operator at least by the standards of Bauke {\textit et al.} 

However, 
the FW spin does not transform covariantly under the Lorentz transformations \cite{Chakarbarti,Good62}. The physical laws including the law of spin dynamics should be Lorentz covariant, hence the FW spin is not suitable to describe the spin dynamics. 
Hence, there still remains a puzzle about what is a proper relativistic spin operator. The confusion has lead difficulties to interpret physical systems such as the Stern-Gerlach deflection of relativistic electron beams \cite{Palmer13,Saldanha,Wen16}, a relativistic Bell test \cite{Bittencourt18} and the experimental results of relativistic electron vortex beams \cite{Birula,Barnett,BirulaComment,BarnettReply}.

Wigner \cite{Wigner1939} classified massive particles using irreducible unitary representations of the Poincar\'e group, characterized by a rest mass $m$ and a spin $s$. The symmetry of our 3+1 dimensional free space-time, called Minkowski space-time, is the Poincar\'e symmetry, which has additional translational symmetry to the homogeneous Lorentz symmetry. In Wigner's classification, spin $s$ labels the irreducible unitary representation of the $SU(2)$ subgroup of the Poincar\'e group, so called the Wigner little group, under which the four-momentum of a given particle is invariant. Therefore, spin operators for massive relativistic particles should be (infinitesimal) generators of the $SU(2)$ little group, which identify an internal angular momentum to generate an internal space symmetry. 

The well-known Pauli-Lubanski (PL) vector was defined to pick out only the internal angular momentum part of the total angular momentum \cite{Bogolubov,Lubanski}. The 4-components of the PL vector with a fixed momentum become generators of the $SU(2)$ little group. 
Then a natural candidate of a spin operator is the PL vector itself with a scale factor $1/m$ to make the PL vector a dimensionless operator in natural unit, which is called spin four-vector in Ref. \cite{Terno2016}. 
However, the PL vector has 4-components other than the three-components of a spin and, as is known, the spatial components of the PL vector in a moving frame of the momentum do not satisfy the usual $su(2)$ algebra \cite{Bogolubov}. This fact implies that spin operators, satisfying the usual $su(2)$ algebra, should be determined from a linear combination of the four-components of the PL vector, which becomes a linear combination of the Lorentz generators at the reference frame of the fixed momentum. 

One of us derived such spin operators from the scratch in the primitive manuscript \cite{Our18}. In this paper we first re-derive the spin operators from a general linear combination of the four-components of the PL vector by imposing two physically natural requirements, in order to make the paper self-contained and logically clear. The two requirements, which would be minimal, are for spin to satisfy the $su(2)$ algebra and to transform covariantly under the restricted Poincar\'e group without time reversal and space inversion. The two requirements derives only two spin operators for the restricted Poincar\'e group, which exchange to each other under parity (space inversion) operation. Hence the space inversion requires that the relativistic-covariant spin of the Poincar\'e group should be the direct sum of the two spin operators, which we call the field spin, because whose representation is composed of the classical fields with mass and spin. We will investigate the Hermiticity of the field spin. 

Bogolubov {\textit et al.} derived other spin, called the Wigner spin in Ref. \cite{Terno2016}, also from a linear combination of the four-components of the PL vector by imposing different conditions \cite{Bogolubov}. 
{We show that the Wigner spin is valid as the generator of the $SU(2)$ little group only at the rest frame of the momentum. This fact suggests that the field spin is the unique spin consistent with the fact that a spin is an internal space rotation operator at an arbitrary frame of momentum. In fact, it will be shown that the FW spin is also reproduced from the field spin. We study the Wigner rotation of the the field spin as a physical requirement for the Lorentz covariance. Wigner rotation has been usually considered as the transformation of the spin state \cite{Wigner1939,WeinbergQFT}, not as the expectation value of the spin operators. Wigner rotation is manifested in the hyperfine structure of atomic spectra due to Thomas precession \cite{Thomas}. Therefore, the expectation value of the physical spin should describe the Wigner rotation properly. It will be shown that the only spin to describe the Wigner rotation properly for the Dirac spinor is the covariant spin derived from the field spin.  

The paper is organized as follows. In Sec. \ref{sec:CSPN} we first re-derive the two spin operators that generate the $SU(2)$ subgroup of the restricted Poincar\'e group from the scratch. Then we construct the field spin operator by the direct sum of the two spin operators and study its Hermiticity. In sec. \ref{sec:WGN} we show that the field spin operator is the unique spin operator having consistent meaning as the generator of the $SU(2)$ little group of the Poincar\'e group obtained from reasonable physical requirements and study the Wigner rotation of the field spin operator as an additional physical requirements for the Lorentz covariance. In sec. \ref{sec:CONC} we conclude with our results.


 \section{Relativistic-covariant spin operator as a physical observable}
\label{sec:CSPN}

\subsection{Derivation of relativistic-covariant spin operator}
\label{sec:DERV}
The laws of physics {in special relativity are invariant} under the following coordinate transformations
\begin{eqnarray}
(\Lambda, a) \,: {x}^{\prime\mu}=\Lambda^{\mu}_{\phantom{\nu}\nu}x^\nu+ a^\mu
\end{eqnarray}
in the Minkowski space-time with the metric tensor $g_{\mu\nu}= \mathrm{diag} (+,-,-,-)$. $a^\mu$ are an arbitrary constant 4-vector and $\Lambda^\mu_{\phantom{\mu}\nu}$ is a Lorentz transformation matrix \cite{WeinbergQFT}. The transformations $(\Lambda,a)$ form the Poincar\'e group.

As was shown by Wigner \cite{Wigner1939}, the spin of a massive elementary particles classifies the irreducible representations of the Wigner's $SU(2)$ little group, which is a subgroup of the Poincar\'e group. This implies that the square of a spin operator should be the second Casimir of the Poincar\'e group, which is proportional to the square of the PL vector \cite{Bogolubov}. The PL vector operator $W^\mu$ is defined as \cite{Lubanski,Bogolubov}
\begin{eqnarray}
\label{eq:PL}
 W^\mu = \frac{1}{2} \epsilon^{\,\mu\nu\rho\sigma} J_{\nu\rho} P_\sigma,
\end{eqnarray}
 where the $4$-dimensional Levi-Civita $\epsilon_{\,\mu\nu\rho\sigma}$ takes the value $+1$ for $\epsilon_{0123}=\epsilon^{1230}$, $J^{\mu\nu}$ are the generators of the $4$-dimensional homogeneous Lorentz group, and $P^\mu$ are the generators of the space-time translation group. 
 We use Einstein summation convention for the Greek indexes $\mu \in \{0,1,2,3\}$ and also for Latin indexes $k\in \{1,2,3\}$.

In this section, we re-derive spin operators in a more compact way than in the primitive manuscript \cite{Our18}. The PL vector operator $W^\mu$ is a generator of the $SU(2)$ little group at a fixed momentum $p^\mu$ because $W^\mu$ in Eq. (\ref{eq:PL}) at the fixed momentum becomes the linear combination of the Lorentz generators $J^{\mu\nu}$ and $[W^\mu,P^\nu]=0$. We omit the word 'operator' freely for simplicity when a context would clarify the usage and we use capital letters for operators and small letters for usual vectors, e.g. $P^\mu$ and $p^\mu$. 

However, as was mentioned in the
Introduction, spin operators could not be identified by the PL vector itself, then a spin operator $S^k$ ($k$-th component of a spin vector ${\bf S}$) as a generator of the $SU(2)$ little group should be determined from the following general linear combination of the 4-components of the PL vector:
\begin{eqnarray}
\label{eq:SCD}
S^k=a_{k,\mu}W^\mu,
\end{eqnarray}
where the coefficients $a_{k,\mu}$ are functions of only the momentum generator $P^\mu$ because $S^k$ should be a linear combination of $J^{\mu\nu}$ at a fixed momentum. 

To probe proper physical requirements for the $S^k$ in Eq. (\ref{eq:SCD}), it is necessary to remind the tensor nature of angular momentum. 
In non-relativistic theory, an orbital angular momentum is defined by the cross product of a position and a momentum $3$-vector. However, in Minkowski space-time, on which special relativity and Poincar\'e symmetry is formulated, a position and a momentum are defined as $4$-vectors. There is no cross product between two 4-vectors \cite{Carroll}.

In Minkowski space-time, a spatial angular momentum is defined as an antisymmetric tensor:
\begin{eqnarray}
L^{\mu\nu}= X^\mu P^\nu -X^\nu P^\mu,
\end{eqnarray}
where $X^\mu$ and $P^\mu$ are 4-position and 4-momentum vector, respectively. A spatial angular momentum $3$-vector $L^k$ is usually defined by the spatial angular momentum tensor as
\begin{eqnarray}
L^{k}=\frac{1}{2}\epsilon_{klm}L^{lm},
\end{eqnarray}
where $\epsilon_{klm}$ is the three-dimensional Levi-Civita with $\epsilon_{123}=1$. 

To specify the covariant transformations under the full Lorentz group, $L^k$ should be considered as $k0$-component of the following Hodge dual of the spatial angular momentum tensor \cite{Carroll}:
\begin{eqnarray}
\ast{L}^{\mu\nu}=\frac{1}{2}\epsilon^{\,\mu\nu\rho\sigma}L_{\rho\sigma}.
\end{eqnarray}
As an angular momentum, spin $S^k$ should be also considered as the $k0$-component of the Hodge dual of the spin tensor $S^{\mu\nu}$: 
\begin{eqnarray}
\label{eq:SPHD}
S^{k}=\ast{S}^{k0}=\frac{1}{2}\epsilon_{klm}S^{lm}.
\end{eqnarray} 
 
Therefore, we impose the following two conditions on $S^k$ to be a spin operator that generates the $SU(2)$ little group: \\
1. $S^k$ should satisfy the usual $su(2)$ algebra (angular momentum condition):
\begin{eqnarray}
\label{eq:AMC}
[S^i,~ S^j] =i\epsilon_{ijk}S^k .
\end{eqnarray} 
2. $S^k$ should transform as a $k0$-component of a second-rank tensor under Lorentz transformations (tensor condition):
\begin{eqnarray}
\label{eq:TCD}
i[J^{\mu\nu}, \ast{S}^{k0}]=g^{\mu k}\ast{S}^{\nu 0}- g^{\nu k}\ast{S}^{\mu 0}- g^{\mu 0}\ast{S}^{\nu k}+g^{\nu 0}\ast{S}^{\mu k} .
\end{eqnarray}
These two conditions would be minimal and are valid to obtain the spin operators of the restricted Poincar\'e group. We will show that the spin operator for the Poincar\'e group (including space inversion) is constructed from the spin operators for the restricted Poincar\'e group. 

Unlike our previous derivation in Ref. \cite{Our18}, where first required that $S^k$ in Eq. (\ref{eq:SCD}) should transform as a $3$-vector under space rotations, we impose tensor condition in Eq. (\ref{eq:TCD}) directly on $S^k$ ($\ast{S}^{k0}$). The derivation procedure is straightforward but tedious, so is briefly explained. 
First we consider the case of $\mu=l$ and $\nu=m$ for the tensor condition in Eq. (\ref{eq:TCD}), then the tensor condition becomes
\begin{eqnarray}
\label{eq:LMTT}
i[J^{lm},~\ast{S}^{k0}]=\delta_{mk}\ast{S}^{l0}-\delta_{lk}\ast{S}^{m0},
\end{eqnarray} 
where $\delta_{mk}$ is Kronecker delta: $1$ for $m=k$ and 0 for others. Eq. (\ref{eq:LMTT}) is rewritten by using $W^\mu$ as
\begin{eqnarray}
\label{eq:RTPL}
 i[J^{lm}, a_{k,\mu}]W^\mu + a_{k,\mu} (g^{\mu l}W^m -g^{\mu m}W^l) = \delta_{mk}a_{l,\mu}W^\mu-\delta_{lk}a_{m,\mu}W^\mu.
\end{eqnarray} 
Eq. (\ref{eq:RTPL}) should be satisfied for arbitrary $W^\mu$'s that are transformed by LT. 
Equating coefficients of $W^\mu$ on both sides of Eq. (\ref{eq:RTPL}), one obtains 
\begin{eqnarray}
a_{k,0}&=& f_0(P^0)P^k, \\ \nonumber
a_{k,k}&=& f_1(P^0)+f_2(P^0)P^kP^k \\ \nonumber
a_{k,m \neq k}&=& f_2(P^0)P^k P^m + f_3(P^0)\epsilon_{kml}P^l, 
\end{eqnarray}
where $f_0(P^0)$, $f_1(P^0)$, $f_2(P^0)$, and $f_3(P^0)$ are functions of $P^0$. 

To fix $a_{k,\mu}$ further, it is enough to consider the case of $\mu=0$ and $\nu=k$:
\begin{eqnarray}
\label{eq:BSTPL}
i[J^{0k},~\ast{S}^{k0}]=0.
\end{eqnarray} 
Equating the coefficients of $W^\mu$ on the left-hand side in Eq. (\ref{eq:BSTPL}) to be zero can fix the functional forms of $f_0(P^0)$, $f_1(P^0)$, $f_2(P^0)$, and $f_3(P^0)$ as
\begin{eqnarray}
f_0(P^0)=a, ~f_1(P^0)= -a P^0, ~f_2(P^0)=0, ~f_3(P^0)=b,
\end{eqnarray}
where $a$ and $b$ are Lorentz invariant constant. Then the spin $S^k$ has the form:
\begin{eqnarray}
\label{eq:SPTP}
S^k=a_{k,\mu}W^\mu= aP^k W^0 -a P^0 W^k + b \epsilon_{kml}P^l W^m
\end{eqnarray}

Lorentz invariant constants $a$ and $b$ are determined as 
\begin{eqnarray}
a=-\frac{1}{m^2} \mbox{ and } b=\pm i\frac{1}{m^2}
\end{eqnarray}
by imposing the angular momentum condition of Eq. (\ref{eq:AMC}) on the $S^k$ in Eq. (\ref{eq:SPTP}). The final results become
\begin{eqnarray}
\label{eq:SPPM}
S^k_{\pm}=  \frac{1}{m^2}\left( P^0 W^k-P^k W^0  \right) \pm i \frac{1}{m^2}\epsilon_{kml}P^l W^m,
\end{eqnarray}
which are the same as the previous result in Ref. \cite{Our18}. 
The two spins $S^k_{\pm}$ become $W^k/m$ at {the rest frame of the momentum}, as expected for the spin at {the rest frame of the momentum}. 
These two spin operators are also the same as those obtained by Ryder \cite{RyderGen} using a commutator of $W^\mu$ and its dual. This fact confirms that the two spin operators $S^k_{\pm}$ are the only covariant spin operators for massive case of the restricted Poincar\'e group.     

\subsection{Spin as physical observable}
\label{sec:SPNPH}

From now on we focus on the massive spin 1/2 case, to which all known fundamental massive fermions belong. 
The two spin operators $S^k_\pm$ in Eq. (\ref{eq:SPPM}), derived for massive case of the restricted Poincar\'e group, exchange to each other by the operation of the parity (space inversion). 
Hence the space inversion in the Poincar\'e group requires that the spin operator $S^k$ should be the direct sum of the two spin operators $S^k_+$ and $S^k_-$, in order to represent the Poincar\'e group, as follows
\begin{eqnarray}
\label{eq:4SP}
S^k = S^k_+ \oplus S^k_-=\frac{1}{m^2}\left( P^0 W^k-P^k W^0  \right) +i \gamma^5\frac{1}{m^2}{\epsilon_{klm}}P^l W^m
\end{eqnarray}
where $\gamma^5$ is defined by $\gamma^5= i\gamma^0\gamma^1\gamma^2\gamma^3$ with usual Dirac $\gamma$-matrices $\gamma^\mu$ \cite{SchwartzQFT}. 
The eigenspinors of $S^3$ become the following two kinds of 4-spinors in the chiral representation \cite{Our18}:
\begin{subequations}
\begin{eqnarray}
\label{eq:P4SP}
u(p,\lambda)&=&\psi_+(p,\lambda)\oplus (\psi_-(p,\lambda))= \left(\begin{array}{c} \psi_+(p,\lambda) \\ \psi_-(p,\lambda) \end{array}\right), \\
\label{eq:AP4SP}
v(p,\lambda)&=&\psi_+(p,\lambda)\oplus (-\psi_-(p,\lambda))= \left(\begin{array}{c} \psi_+(p,\lambda) \\ -\psi_-(p,\lambda) \end{array}\right),
\end{eqnarray}
\end{subequations}
where $p=(E_{\bf p}, {\bf p})$ and $E_{\bf p}=\sqrt{{\bf p}\cdot{\bf p}+m^2}$. $\psi_+(p,\lambda)$ and $\psi_-(p,\lambda)$ are the left-handed and the right-handed spinors satisfying the following eigenvalue equations 
\begin{subequations}
\begin{eqnarray}
\label{eq:EVELSP}
S^3_+ \psi_+(p,\lambda) &=& \lambda \psi_+(p,\lambda) \\
\label{eq:EVERSP}
 S^3_- \psi_-(p,\lambda) &=& \lambda \psi_-(p,\lambda)
\end{eqnarray}
\end{subequations}
with $\lambda=\pm 1/2$. 

Then the representation space of the Poincar\'e group, called the covariant (chiral) Dirac representation, consists of the Dirac spinor fields given by the superposition of the following Dirac plane wave spinors for massive particle and antiparticle:
\begin{eqnarray}
\label{eq:DPWSP}
\Psi_P(x;p,\lambda) =e^{-ip \cdot x}u(p,\lambda) \mbox{ and } \Psi_{AP}(x;p,\lambda) =e^{ip \cdot x}v(p,\lambda),
\end{eqnarray}
where $p\cdot x=p^\mu x_\mu$. The subscripts $P$ and $AP$ denote particle and antiparticle, respectively. The Dirac spinor field is definitely an eigenstate of the spin $S^3$. In this sense we call $S^k$ the field spin.

The field spin operator $S^k$ in and of itself {seems to be not Hermitian due to $i$ in the last term of Eq. (\ref{eq:4SP}). 
However, Hermiticity of the operator should be determined by the actions of the operator on the state, which define the operator.} 
The (infinite-dimensional) matrix elements, $(q, \lambda';p,\lambda)$, of the field spin $S^3$ is defined as
\begin{eqnarray}
S^3_{q,\lambda';p,\lambda} = \int d^3 {\bf x} \psi^\dagger(x;q,\lambda') S^3 \psi(x;p,\lambda),
\end{eqnarray}
which becomes
\begin{eqnarray}
S^3_{q,\lambda';p,\lambda} &=& \lambda (2\pi)^3 \delta({\bf p}-{\bf q})\delta_{\lambda\lambda'}
\end{eqnarray}
by using $u^\dagger(p,\lambda)v(\bar{p},\lambda)=0$ with $\bar{p}=(E_{\bf p}, -{\bf p})$, where $\psi(x;p,\lambda)$ and $\psi(x;q,\lambda')$ are the normalized superpositions of the corresponding Dirac particle and antiparticle plane wave spinors in Eq. (\ref{eq:DPWSP}). 
Then the field spin $S^3$ satisfies the Hermiticity condition as follows
\begin{eqnarray}
S^3_{q,\lambda';p,\lambda} &=& \int d^3 {\bf x} \left (S^3 \psi(x;q,\lambda') \right)^\dagger \psi(x;p,\lambda)  \\ \nonumber
&=&(S^3_{p,\lambda;q,\lambda'})^*.
\end{eqnarray}
The field spin $S^k$ was shown to be a constant of motion by using the Noether theorem in Ref. \cite{Our18}. Hence the field spin $S^k$ satisfies the basic properties as a physical observable. In the next section we show that the field spin $S^k$ is the unique spin operator generating the little group transformation of the Poincar\'e group properly and study the Wigner rotation as an additional physical requirement. 



\section{Uniqueness and Wigner Rotation}
\label{sec:WGN}

\subsection{Uniqueness}

Only one other spin operator, which could be derived from the generators of the Poincar\'e group satisfying reasonable physical requirements, was known by Bogolubov {\textit et al.} \cite{Bogolubov}:  
\begin{eqnarray}
\label{eq:WS}
S^k_W= \frac{1}{m}\left( W^k - \frac{W^0 P^k}{m+P^0}\right),
\end{eqnarray} 
called the Wigner spin operator in Ref. \cite{Terno2016}. 
Both the field spin $S^k$ in Eq. (\ref{eq:4SP}) and the Wigner spin in Eq. (\ref{eq:WS}) are expressed in terms of the PL vector. The PL vector in Eq. (\ref{eq:PL}) can be rewritten by using the internal angular momentum tensor $S^{\mu\nu}$ as
\begin{eqnarray}
\label{eq:PLSP}
W^\mu = \frac{1}{2}\epsilon^{\mu\nu\rho\sigma}S_{\nu\rho}P_\sigma,
\end{eqnarray}
because the orbital part of $J_{\nu\rho}$ in $W^\mu$ vanishes, because $\epsilon^{\mu\nu\rho\sigma} (X_\nu P_\rho -X_\rho P_\nu)P_\sigma=0$. 
Therefore, both the two spin operators $S^k$ and $S^k_W$ in Eqs. (\ref{eq:4SP}) and (\ref{eq:WS}) as the generator of the internal $SU(2)$ little group should be consistent with the internal angular momentum tensor $S^{\mu\nu}$ in Eq. (\ref{eq:PLSP}).  

We check the consistency of the field spin $S^k$ and the Wigner spin $S^k_W$ by using the covariant and the non-covariant $W^\mu$ operators satisfying $W^\mu_R=(0, m\boldsymbol{\Sigma}/2)$ at the rest frame of the momentum. ${\Sigma}^k$, the $k$-component of $\boldsymbol{\Sigma}$, is defined by the direct sum of the usual Pauli matrices as $\sigma \oplus \sigma^k$. First, we consider the covariant $W^\mu$ that is obtained as follows
\begin{eqnarray}
\label{eq:CPL}
W^0 = \frac{\boldsymbol{\Sigma} \cdot \mathbf{p}}{2} 
\mbox{ and  } W^i = m \frac{\Sigma^i}{2} + \frac{\boldsymbol{\Sigma} \cdot \mathbf{p} p^i }{2(m+E_{\bf p})}
\end{eqnarray} 
at the frame of the momentum ${\bf p}$ by using $W^\mu = {\mathcal{L}_p}^\mu_{\phantom{\mu}\nu}W^\nu_R$, where $\mathcal{L}_p$ is the standard Lorentz transformation such that $p^\mu={\mathcal{L}_p}^\mu_{\phantom{\nu}\nu} k^\nu$ for $k^\nu=(m, {\bf 0})$. Here $\mbox{\boldmath $\Sigma$} \cdot \mathbf{p}=\Sigma^k p^k$.
Then the field spin $S^k$ in Eq. (\ref{eq:4SP}) by using $W^\mu$ in Eq. (\ref{eq:CPL})  becomes 
\begin{eqnarray}
\label{eq:DCSP}
S^k=U(\mathcal{L}_{ p}) \frac{\Sigma^k}{2}U^{-1}(\mathcal{L}_{ p}),
\end{eqnarray}
where
\begin{eqnarray}
U(\mathcal{L}_p)=e^{-\gamma^5 \boldsymbol{\Sigma}\cdot{\boldsymbol{\xi}}/2}=\frac{E_{\bf p}+m -\gamma^5 \boldsymbol{\Sigma}\cdot{\bf p}}{\sqrt{2m(E_{\bf p}+m)}}
\end{eqnarray} 
is the spinor representation of $\mathcal{L}_p$ and $ {\boldsymbol{\xi}}$ is the rapidity with ${\boldsymbol{\xi}}=2 \hat{\bf p}\tanh^{-1}\frac{|{\bf p}|}{E_{\bf p}}$ and $|{\bf p}|=\sqrt{{\bf p}\cdot {\bf p}}$. The field spin $S^k$ in Eq. (\ref{eq:DCSP}) is the covariant spin defined in the covariant Dirac representation, which satisfies
\begin{eqnarray}
\label{eq:CSP}
S^k=\ast{S}^{k0}={\mathcal{L}_p}^k_{\phantom{\nu}\mu} {\mathcal{L}_p}^0_{\phantom{\nu}\nu} \frac{\Sigma^{\mu\nu}}{2},
\end{eqnarray} 
where 
\begin{eqnarray}
\Sigma^{\mu\nu}=\frac{i}{4}[\gamma^\mu,\gamma^\nu].
\end{eqnarray}
 One can easily check the eigenvalue equation for $S^3$ in Eq. (\ref{eq:DCSP}), because $u(p,\lambda)$ and $v(p,\lambda)$ in Eqs. (\ref{eq:P4SP}) and (\ref{eq:AP4SP}) are obtained by the standard Lorentz transformation from the rest spinors: 
\begin{eqnarray}
\label{eq:DCR}
u(p,\lambda)=U(\mathcal{L}_p) u(k,\lambda) \mbox{ and } v(p,\lambda)=U(\mathcal{L}_p) v(k,\lambda)
\end{eqnarray}
where $u(k, +1/2)= (1,0,1,0)^T$, $u(k, -1/2)= (0,1,0,1)^T$, $v(k, +1/2)= (1,0,-1,0)^T$, $u(k, +1/2)= (0,1,0,-1)^T$, and $T$ denotes transpose. 

On the other hand, the Wigner spin $S^k_W$ given by the covariant $W^\mu$ in Eq. (\ref{eq:CPL}) becomes 
\begin{eqnarray}
\label{eq:CWS}
S^k_W=\frac{1}{2}\Sigma^k,
\end{eqnarray}
which is the same as the rest frame spin, not the expected covariant spin in Eq. (\ref{eq:CSP}). Hence the consistency check for the covariant $W^\mu$ implies that the Wigner spin $S^k_W$ is valid only at the rest frame of the momentum, but the field spin is valid at an arbitrary frame of the momentum ${\bf p}$.


Next we investigate the consistency of the two spins by using the non-covariant $W^\mu$, which is defined from the non-covariant internal angular momentum tensor 
\begin{eqnarray}
\label{eq:SPTNC}
S^{\mu\nu}=  \frac{\Sigma^{\mu\nu}}{2}
\end{eqnarray}
at an arbitrary frame of the momentum ${\bf p}$. The non-covariant $W^\mu$ becomes
\begin{eqnarray}
\label{eq:NCPL}
W^0 = \frac{\boldsymbol{\Sigma} \cdot \mathbf{p}}{2} 
\mbox{ and  } W^i = \frac{1}{2} \, p^0 \,\Sigma^k + \frac{i}{2} \left(\gamma^0 \boldsymbol{\gamma}\times {\bf p} \right)^k,
\end{eqnarray}
where $\times$ is the three-dimensional cross product. The angular momentum tensor $\Sigma^{\mu\nu}/2$ produces the 4-dimensional Pauli spin 3-vector $\Sigma^k/2$, hence the consistency requires that the spin determined by the non-covariant $W^\mu$ in Eq. (\ref{eq:NCPL}) should be the 4-dimensional Pauli spin 3-vector $\Sigma^k/2$. The direct calculations give the following results
\begin{subequations}
\begin{eqnarray}
\label{eq:CSTFSP}
S^k &=& \frac{\Sigma^k}{2} \\
\label{eq:CSTWSP}
S^k_W &=& \frac{i}{2m}\gamma^5 (\boldsymbol{\Sigma}\times {\bf P})^k +\frac{1}{2m}\Sigma^k P^0 -\frac{1}{2m(m+P^0)} (\boldsymbol{\Sigma}\cdot{\bf P})P^k. 
\end{eqnarray}
\end{subequations}
The Wigner spin $S^k_W$ in Eq. (\ref{eq:CSTWSP}) can be $\Sigma^k/2$ only at the rest frame of the momentum, that is, is only valid at the rest frame of the momentum. This implies that the Wigner spin is not the FW spin that is defined as $\Sigma^k/2$ at the frame of the momentum ${\bf p}$ in the FW representation, which has been accepted by many authors. 

On the other hand, Eq. (\ref{eq:CSTFSP}) shows that the definition of the field spin $S^k$ in Eq. (\ref{eq:4SP}) is consistent to let $S^{\mu\nu}=\Sigma^{\mu\nu}/2$ at the frame of the momentum ${\bf p}$. The spin representation of $\Sigma^k/2$ at the frame of an arbitrary momentum ${\bf p}$ is known as the FW spin in the FW representation \cite{FW}, however the momentum for the negative energy is not covariant in the FW representation. Hence the field spin $S^k=\Sigma^k/2$ in Eq. (\ref{eq:CSTFSP}) is the FW spin only for the positive energy (particle) state and valid in the particle and the antiparticle representation space generated by the constant 4-dimensional spinors like $u(k,\pm 1/2)$ and $v(k, \pm 1/2)$. This implies that the field spin $S^k$ in Eq. (\ref{eq:4SP}) is the unique spin having the consistent meaning of the internal $SU(2)$ little group generator and valid for the 4-spinor fields in all representations of the Poincar\'e group.

\subsection{Wigner Rotation}

The unitary representation of the Poincar\'e group shows a rotation under a general Lorentz transformations \cite{Wigner1939,WeinbergQFT}. This is a kinematic consequence from the fact that the successive application of two non-collinear Lorentz boosts is not a pure Lorentz boost, but a pure Lorentz boost combined with a spatial rotation, known as the Wigner rotation. Thomas precession is a manifestation of the Wigner rotation, which explains the hyperfine structure of atomic spectra \cite{Thomas}. The most arguments of the Wigner rotation in the unitary representation of the Poincar\'e group only used the transformations of the state, however, the physical interpretation of the Wigner representation requires the explicit information of the direction of the spin \cite{VedralWR}. We will investigate whether the field spin and the Dirac spinor providing the covariant Dirac representation space describe the Wigner rotation properly. 


Let us consider a general Lorentz transformation $\Lambda$ that transforms $p^\mu$ to $q^\mu=\Lambda^\mu_{\phantom{\mu}\nu}p^\nu$ or $q=\Lambda p$ in a shorthand notation. The Dirac spinors also show the similar relations to the unitary representation states under the Lorentz transformation $U(\Lambda)$ as follows  
\begin{subequations}
\begin{eqnarray}
\label{eq:GLTSPRF}
U(\Lambda)\psi(p,\lambda)&=&U(\mathcal{L}_{\Lambda { p}})U(\mathcal{L}^{-1}_{\Lambda p} \,\Lambda \,\mathcal{L}_p)\psi({ k},\lambda) \\
\label{eq:GLTSPAF}
&=&\sum_{\lambda'} \mathcal{D}_{\lambda' \lambda}(R)\psi({ q},\lambda'),
\end{eqnarray}
\end{subequations}
because $R=\mathcal{L}^{-1}(\Lambda p) \,\Lambda\, \mathcal{L}_p$ is the Wigner rotation, 
where $\psi(p,\lambda)$ is either $u(p,\lambda)$ or $v(p,\lambda)$. $ \mathcal{D}_{\lambda' \lambda}(R)$ is the 4-dimensional block diagonal matrix made of two identical conventional 2-dimensional matrix representing $SU(2)$ group, because the Dirac spinor is the direct sum of two 2-dimensional spinors with the same spin index.

The field spin $S^k$ can describe the Wigner rotation of the spin properly via Eqs. (\ref{eq:GLTSPRF}) and (\ref{eq:GLTSPAF}) under two assumptions. One is that there is a proper spinor representation $U(\Lambda)$ isomorphic to a general Lorentz transformation $\Lambda$. The other is that the Dirac spinor $\psi(q, \lambda)$ is the eigenstate of the field spin ${S}^3$. 
Then Eq. (\ref{eq:GLTSPAF}) describes the Wigner rotation of the spin through the expectation value at the reference frame of the momentum ${\bf q}$ as follows
\begin{eqnarray}
\left(\sum_{\lambda^{\prime\prime}} \mathcal{D}_{\lambda^{\prime\prime} \lambda}(R)\psi({ q},\lambda^{\prime\prime}) \right)^\dagger S^l \sum_{\lambda'} \mathcal{D}_{\lambda' \lambda}(R)\psi({ q},\lambda')=R^l_{\phantom{l}m}\left< S^m \right>,
\end{eqnarray}
where $\langle S^m \rangle =\psi^\dagger({ q},\lambda) S^m \psi(q,\lambda)$.
The field spin $S^3$ satisfies the eigenvalue equation:
\begin{eqnarray}
\label{eq:EGVE}
S^3 \psi({ q},\lambda)=\lambda \psi({ q},\lambda).
\end{eqnarray}
The spinor representation $U(\Lambda)$ is provided by the element of a complexified $SU(2)$ group through the field spin ${\bf S}$ as follows \cite{Our18} 
\begin{eqnarray}
\label{eq:GLTSR}
U(\Lambda)= e^{i{\bf S}\cdot\boldsymbol{\theta}{-}\gamma^5 {\bf S}\cdot\boldsymbol{\xi}},
\end{eqnarray}
where the rotation angle $\boldsymbol{\theta}$ and the rapidity $\boldsymbol{\xi}$ are determined by $\Lambda$. Moreover, the field spin $S^k$ is the only spin to describe the Wigner rotation properly for the Dirac spinor $\psi(p,\lambda)$, because the Wigner rotation is clearly defined as the usual rotation in the rest frame of the momentum and the field spin $S^k$ in Eq. (\ref{eq:DCSP}) is the sole representation to satisfy the eigenvalue equation equivalent to the eigenvalue equation in the rest frame of the momentum. 

\section{Conclusions}
\label{sec:CONC}

 In this paper, we derived the relativistic-covariant spin operator for relativistic massive fields from the point of view of space-time symmetry in special relativity and investigated its uniqueness and physical properties. We first re-derived the two spin operators, which generate the $SU(2)$ little group of the restricted Poincar\'e group, from a general linear combination of the four-components of the PL vector. In the derivation we imposed two natural requirements, which are for spin to satisfy the $su(2)$ algebra and to transform covariantly under the restricted Lorentz group. 
The derived two spin operators exchange to each other by space inversion, hence they cannot be a spin operator of the Poincar\'e group including space inversion. 

It was shown that the physically observable spin, the field spin, comes from the Poincar\'e symmetry with an additional space inversion symmetry. The field spin operator is the direct sum of the two spin operators for the restricted Poincar\'e group, which gives the direct sum space as the representation space of the Poincar\'e group. The representation space in the covariant represrentation derived by the field spin becomes the well-known covariant Dirac spinor field space for massive spin $s=1/2$ case, which has a particle and an antiparticle subspace. We have shown that the field spin operator is Hermitian by direct calculation of the infinite-dimensional matrix elements of the field spin given by the action of the field spin on the Dirac spinors, even though the form of the field spin operator in and of itself is not Hermitian. The field spin is also a constant of motion for free Dirac spinor field. Hence the field spin satisfies the necessary condition as a physical observable. 


It was also shown that the field spin is the unique spin operator to generate the internal space rotation and to describe the Wigner rotation for the Dirac spinor field properly. In summary, the field spin is supposed to be a physical spin that describes the spin of a particle and an antiparticle in relativistic quantum mechanics, because the field spin is the unique spin to be directly derived from the Poincar\'e symmetry, which is valid in all representations and to have all the desirable properties as a physical spin. Therefore, we think that the puzzle what is a proper relativistic spin operator is solved.

\section*{Acknowledgements}

 This work was supported by the Basic Science Research Program through the National Research Foundation of
Korea (NRF) funded by the Ministry of Education (2019-0300) and by a research grant from Seoul Women’s University(2022-0165).

\end{document}